\documentclass[5p]{elsarticle}

\usepackage{amssymb}
\usepackage{amsthm}
\usepackage{amsmath}
\usepackage{subfig}
\usepackage{multirow}
\usepackage{ulem}
\usepackage{overpic}
\usepackage{soul}

\usepackage{verbatim}
\usepackage{color}

\begin{document}

\begin{frontmatter}

\author{Joel T. Clemmer\corref{cor1}\fnref{SNL}} 
\ead{jtclemm@sandia.gov}
\cortext[cor1]{Corresponding author}
\author{Jeremy B. Lechman\fnref{SNL}}

\affiliation[SNL]{organization={Sandia National Laboratories},
            city={Albuquerque},
            postcode={87185}, 
            state={New Mexico},
            country={USA}}

\title{Onset and impact of plastic deformation in granular compaction}

\begin{abstract}
The role of plastic deformation in the high-pressure compaction of granular material is investigated using bonded particle model simulations.
Grains are discretized into a set of computational particles connected by pairwise bonds. 
Bonds are harmonic up to a plastic onset strain $\epsilon_p$ above which they yield, capping out at a maximum force and producing an elastic-perfectly-plastic-like mechanical response in grains.
Packings containing over one thousand monodisperse spherical grains are isotropically compacted to different packing fractions to quantify how decreasing $\epsilon_p$ softens the rise in pressure and impacts effective elastic properties of the confined system.
By isolating the relative decrease in pressure due to plasticity, we find data can be collapsed across a wide range of values of $\epsilon_p$ suggesting that relatively simple mathematical descriptions may capture plasticity's effect in granular compaction.
Lastly, we study the microscopic statistics of local strains in grains and connect their evolution to the observed macroscopic behavior.
\end{abstract}

\end{frontmatter}

\section{Introduction}

With increasing pressure, granular matter densifies. 
This self-evident phenomenon is at the core of many industrial processes including, but not limited to, the compaction of soil, pharmaceutical tableting, and electrode calendaring in battery manufacturing.
Of course, the final state of a compacted powder, including distributions of residual pore space and material damage, is quite complicated and has significant consequences in the performance of the final product.
This performance may include the mechanical strength of a pill \cite{Mohammed2005}, where defective tablets lead to unnecessary costs, or the nature of a deflagration-to-detonation transition in energetic materials \cite{Baer1986}.
Such applications have motivated a need to quantitatively study and understand granular behavior at high pressures \cite{Ribiere2005, Bares2022}.
In this work, we focus specifically on the role of irreversible plastic deformation, a key mechanism in the densification of many metallic, polymeric, and molecular crystalline feedstock powders, and systematically study its impact on compaction, linking the activation of plasticity on sub-granular length scales to the macroscopic mechanical properties of the compacted powder.

While significant work has gone into studying the compaction of powders, relatively few studies exist that specifically attempt to disentangle the effects of plastic deformation from other physical processes.
In particular, experiments by \citet{Vu2019} compared quasi-2D packings of cylindrical particles consisting of either hyperelastic silicone or elastic-plastic agar hydrogel. 
Their controlled comparison directly demonstrated how significantly plasticity can blunt the rise in pressure at large densities.
Their innovative measurements of granular shape and strain also revealed significant differences in the structure of packings between the two materials.
Using the Material Point Method, \citet{Nezamabadi2021, Nezamabadi2021b} found qualitatively similar behavior in 2D simulations of elastic and elastic-plastic materials compacted to nearly fully dense states.
By controlling a hardening modulus in a bilinear constituitive model, they quantified how a drop in stiffness above plastic yield decreased the pressure required to pack to a given density and proposed a corresponding analytic model to capture their observations.

In this work, we aim to extend such explorations by explicitly focusing on the transition from elastic to plastic behavior, systematically controlling a local plastic yield criterion $\epsilon_p$, in 3D packings of spherical particles.
Based on prior studies \cite{Clemmer2023, Clemmer2024}, we use a bonded particle model to simulate the isotropic compaction of 1,024 elastic-plastic grains described in Sec. \ref{sec:methods}.
First, we consider the macroscopic behavior of systems with different $\epsilon_p$ in Sec. \ref{sec:results_macro}.
At small packing fractions, the pressure of all systems follow the elastic limit before deviating at a $\epsilon_p$-dependent packing fraction as they soften.
Correspondingly, the bulk modulus of the granular packing weakens with decreasing $\epsilon_p$ above this threshold while the shear modulus has minimal dependence on $\epsilon_p$.
Isolating the transition in these datasets in Sec. \ref{sec:results_deviation}, we find that data can be collapsed using a non-trivial power of $\epsilon_p$ and that the reduction in relative pressure grows approximately as a square-root of distance from this transition.
Lastly in Sec. \ref{sec:results_micro}, we show how these observations correspond to the microscopic emergence of plastic deformation in bonds and connect it to distributions of local strain within the system.

\section{Methods}
\label{sec:methods}

Many different numerical approaches to simulate elastic-plastic granular material at a wide range of length scales have been proposed in the literature.
At the largest length scales, simulations treat granular material as a continuum and, with effective parameterization, can accurately capture the entirety of large industrial processes \cite{Dunatunga2015, VanderHaven2024}.
The effectiveness of this approach, of course, relies on choosing an appropriate numerical/constitutive model to represent the underlying granular material. 
Going down in length scales, the discrete element method (DEM) resolves the actual dynamics of individual grains, each represented by a computational particle.
Models for inter-particle contact forces vary in accuracy, sophistication, and computation costs, spanning from purely pairwise interactions \cite{Brake2012} to complex multibody constructions that resolve secondary contacts at high pressures \cite{Zunker2024, Zunker2024b}.
Finally at sub-granular length scales, there are more detailed methods that explicitly resolve the internal elastic and plastic deformation of grains that produces contact forces.
These include, but are not limited to, the material point method (MPM) \cite{Nezamabadi2021, Nezamabadi2021b, Nezamabadi2024} which notably was able to simulate 3D plastic sphere impacts in work by \citet{Saifoori2024}, the multi-particle finite element method \cite{Procopio2005, Harthong2012, Ku2023}, peridynamics \cite{Silling2021}, and bonded particle models (BPM) \cite{Nguyen2021, Clemmer2023}.

Typically, BPMs are relatively simple mesh-free method for simulating solid mechanics \cite{Lisjak2014}.
Grains are subdivided into many computational particles\footnote{In this article, the term particle is exclusively used to refer to the fundamental computational unit not an individual grain of the granular material.} whose positions are numerically integrated based on Newtonian mechanics.
Forces typically include intra-granular attractive/repulsive bonds and inter-granular repulsive contact forces.
To efficiently simulate large systems at long times, we use the parallelized, open-source BPM package developed in the widely used LAMMPS codebase \cite{Thompson2021, Clemmer2024}.

Unlike other mesh-free solid mechanics solvers like MPM or peridynamics, one generally does not model a given stress-strain constitutive equation using a BPM, although exceptions exist \cite{Zhang2019, Andre2023}.
Instead, one must typically construct interparticle forces that produce the desired macroscopic solid mechanics.
This can be easily accomplished for simple mechanical models, like isotropic linear elasticity, but is generally a challenge for more complex models.
However, this limitation is not always necessarily a disadvantage as an appropriate constitutive equation is not guaranteed to be known/available for many granular materials and the need to exactly reproduce a specific response is not essential for theoretical studies of general trends, like the role of plasticity.

The simulations studied in this work are based on those in Ref. \cite{Clemmer2024}, which includes additional description of the methods and context for various choices.
We do not use the multibody interaction described in Ref. \cite{Clemmer2024} which provided a separate knob to adjust Poisson's ratio, but instead introduce plastic deformation into bonds.
Spherical grains with radius $10 d$ are first constructed out of $5,137$ particles of mass $m$ and diameter $d$.
The spherical shape is cut from a larger amorphous packing of particles before smoothing its surface with a repulsive spherical wall to reduce corrugation and the effective friction strength due to geometric interlocking of particles (the smoother model in Ref. \cite{Clemmer2024}). 
Bonds are finally generated between all pairs of particles within a distance of $1.5d$.
Since bonded sets of particles can already represent rotation, rotational degrees of freedom of individual particles are not resolved.

\begin{figure}
\begin{centering}
	\includegraphics[width=0.9\columnwidth]{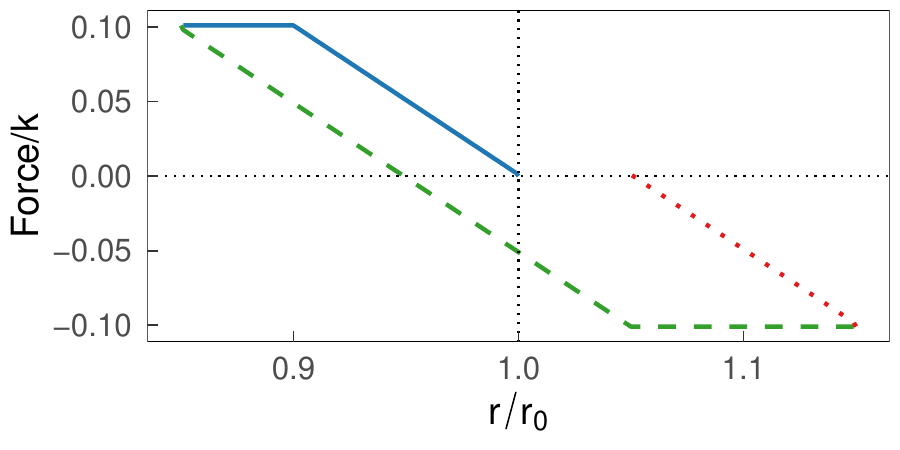}
	\caption{The stiffness-normalized force of a bond that first undergoes compression (solid blue), then extension (dashed green), and finally compression (dotted red) as a function of stretch for $\epsilon_p = 0.1$ where $r_0$ is the initial bond length.}
	\label{fig:bforce}
\end{centering}
\end{figure}

Bonds exert purely pairwise, central-body forces with a magnitude
\begin{equation}
- k (r_\mathrm{eq} - r) - \Gamma \hat{r} \cdot \delta \vec{v}
\label{eq:force_bond}
\end{equation}
where $\vec{r}$ is the displacement between the two bonded particles, $k$ is a stiffness, $\Gamma = \sqrt{km}$ is a damping strength, and $\delta \vec{v}$ is the velocity difference. 
The second term is a simple damping force while the first term is a harmonic spring that can plastic yield. 
Each bond individually calculates its equilibrium length $r_\mathrm{eq}$ based on its own unique strain history \cite{Clemmer2023}. 
At the start of a simulation, $r_\mathrm{eq}$ is equal to the initial distance between the two particles.
However, if the bond ever stretches beyond a plastic activation strain $\epsilon_p$ in either compression or tension, the bond becomes plastically active and $r_\mathrm{eq}$ starts evolving.
Every timestep, $r_\mathrm{eq}$ is adjusted, if necessary, to ensure $|(r - r_\mathrm{eq})/r_\mathrm{eq}|$ never exceeds $\epsilon_p$.
This produces the hysteresis demonstrated in Fig. \ref{fig:bforce} and leads to macroscopic plastic deformation in bulk samples as demonstrated in \ref{sec:bulk}.
The bulk macroscopic behavior is approximately linear-elastic-perfectly-plastic up to a strain of approximately $0.35$ with only minor stiffening (around $10\%$) before substantial deviations emerge at even larger strains as particles begin to accumulate significant non-affine displacement.
Crucially, however, decreasing $\epsilon_p$ adjusts when the the solid yields and enhances the plastic softening during compression.

\begin{figure*}
\begin{centering}
	\includegraphics[width=\textwidth]{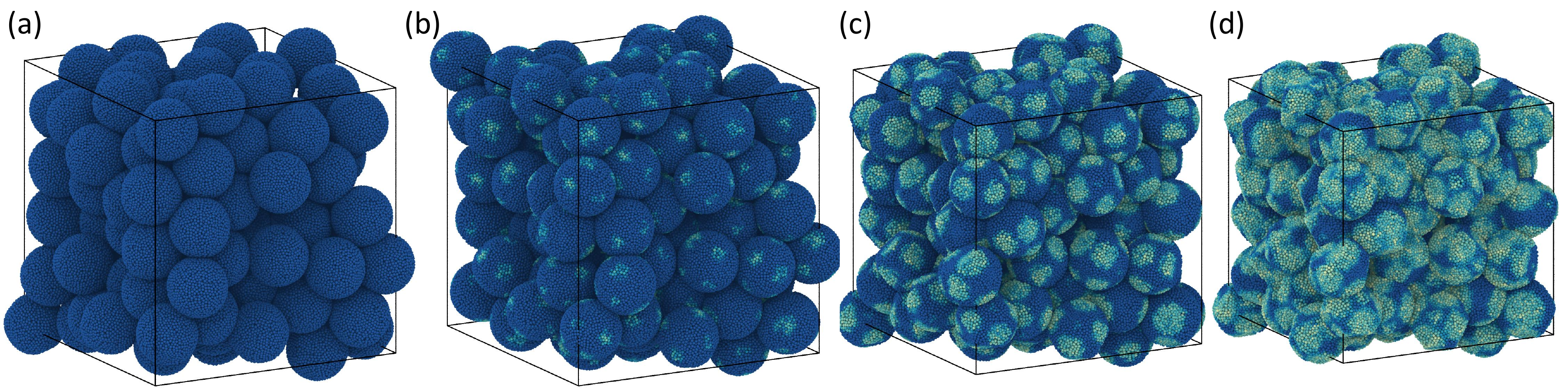}
	\caption{Example small systems with 128 grains compressed to nearly equivalent pressures of $P/E_\mathrm{eff} = (2 \pm 0.1) \times 10^{-2}$ for $\epsilon_p$ = (a) 1.0, (b) 0.05, (c) 0.02, and (d) 0.01 and packing fractions $\phi \approx 0.72$, $0.74$, $0.82$, and $0.91$, respectively. Particle color corresponds to the average excess strain above yield in their associated bonds. Blue represents zero strain above $\epsilon_p$ (no plastic activation) and white represents strains of $\sim 0.25$ over $\epsilon_p$ (significant local plasticity).}
	\label{fig:grain_geom}
\end{centering}	
\end{figure*}

Between two non-bonded particles within a distance of $d$, i.e. surface particles on two contacting grains, a repulsive pairwise, central-body contact force is applied.
The magnitude of this force is
\begin{equation}
k \left(d - r \right) + 50 \frac{k}{d^2} \left(d - r\right)^3 - \Gamma \hat{r} \cdot \delta \vec{v} \ \ .
\label{eq:pairforce}
\end{equation}
The anharmonic term is added to prevent overlaps from approaching $\sim d$ even at very high pressures.

Using a jammed DEM packing as a reference, systems are initialized by placing 1,024 grains in a cubic simulation box that is expanded slightly to break all granular contacts and sit on the verge of jamming.
Using a range of values of $\epsilon_p$ from $0.01$ to $1.0$, systems are then isotropically compacted to a nearly fully dense state, as demonstrated in Fig. \ref{fig:grain_geom}.
Initially, the true strain rate is very small, $10^{-9} \tau^{-1}$, but is regularly increased until reaching $10^{-6} \tau^{-1}$ where $\tau \equiv \sqrt{m/k}$ is the unit of time and simulations use a timestep of $0.1 \tau$.
This allows simulations to explore states that are both very close to jamming and nearly fully dense without running for excessively long times.
No significant changes were detected when running a purely elastic system at much slower rates, indicating simulations are representative of quasi-static compaction.
Pressure is calculated as the sum of virial and kinetic contributions, the latter of which is negligible at these small strain rates.

\section{Results and discussion}

\subsection{Macroscopic compaction response}
\label{sec:results_macro}

\begin{figure}
\begin{centering}
	\includegraphics[width=0.9\columnwidth]{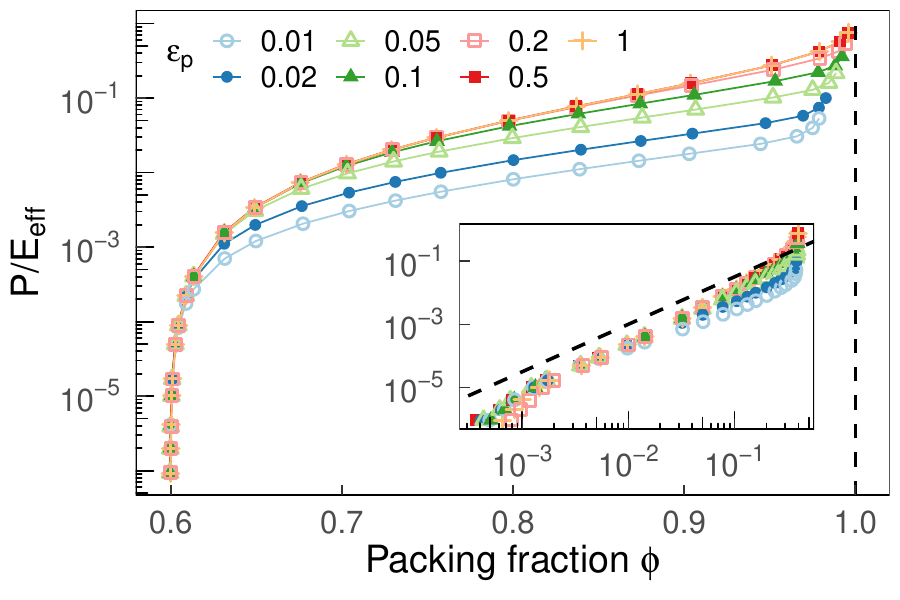}
	\caption{Normalized pressure $P$ as a function of packing fraction $\phi$ at the indicated values of $\epsilon_p$. A dashed vertical line marks a fully dense system. In the inset, data is plotted vs. $\phi - \phi_c$ with $\phi_c = 0.5992$. The dashed line represents a power law with exponent $3/2$.}
	\label{fig:Pcurves}
\end{centering}
\end{figure}

To begin, we consider how the total pressure of the system increases during densification (Fig. \ref{fig:Pcurves}).
Throughout the text, pressures are normalized by the effective modulus of the Hertz force, $E_\mathrm{eff} \equiv 3 K (1- 2 \nu)/ (2 - 2 \nu^2)$ where $K \approx 1.24 k/d$ is the bulk modulus of the solid material constituting grains (calibrated in Ref. \cite{Clemmer2024}) and $\nu = 1/4$ is the Poisson's ratio.
Since grains are deformable and are made up of many potentially overlapping particles, they do not have a well-defined geometric shape.
Therefore, the packing fraction $\phi$ is estimated using Monte Carlo integration where $10^7$ random points are sampled in the simulation cell and checked for collisions with individual particles (with diameter $d$) and the convex hull of each grain \cite{Clemmer2024}.
This approach both avoids double counting overlapping regions between particles which would overestimate $\phi$ \cite{Boromand2019} and includes contributions from the excluded volume within a solid grain consisting of internal gaps between particles.

We start by summarizing the fully-elastic scenario with no plastic deformation, which has already been studied in significant detail by \citet{Cardenas-Barrantes2022} using the non-smooth contact dynamics method and in our earlier work using a BPM \cite{Clemmer2024}.
In Fig. \ref{fig:Pcurves}, the elastic limit is represented by $\epsilon_p = 1.0$, an activation threshold sufficiently high that bonds never plastically activate.
The system first jams at $\phi \approx 0.6$ and then continues compacting until $\phi \sim 0.995$ causing an almost six decade jump in pressure. 
At small $\phi$ near jamming, data is consistent with the expected growth in the mean pressure $P$ as $(\phi - \phi_c)^{3/2}$ \cite{OHern2003} where $\phi_c = 0.5992$ is the estimated packing fraction at jamming, roughly chosen to maximize this power-law domain (Fig. \ref{fig:Pcurves} inset).
Extending the power law to smaller $\phi$ may require larger systems \cite{Goodrich2012} or more carefully jammed states at lower pressures \cite{Agnolin2007}.
At large $\phi$, this power law notably breaks down around $\phi - \phi_c \sim 0.1$ as the rise in pressure accelerates as $\phi$ approaches 1.0.

Now considering other values of $\epsilon_p$, there is initially no detectable dependence on $\epsilon_p$ near jamming as no bonds have plastically activated and simulations are effectively identical.
However, as $\phi$ increases a few percent, $P$ begins to deviate at the smallest value of $\epsilon_p = 0.01$.
As $\phi$ continues to increase, curves systematically peel away from the elastic limit at larger $\epsilon_p$.
The magnitude of this effect on the $P$ can be quite significant, up to a factor of ten at comparable $\phi$ or alternatively up to a $20\%$ difference in $\phi$ at comparable $P$.
This difference is clearly visible in Fig. \ref{fig:grain_geom}.
In the elastic limit, Fig. \ref{fig:grain_geom}(a), the pressure is relatively small and one cannot identify any obvious deformation of grains. 
In comparison at $\epsilon_p = 0.01$ in Fig. \ref{fig:grain_geom}(d), significant flattening of contacts occurs due to plastic softening despite being at equivalent pressures.

\begin{figure}
\begin{centering}
	\includegraphics[width=0.9\columnwidth]{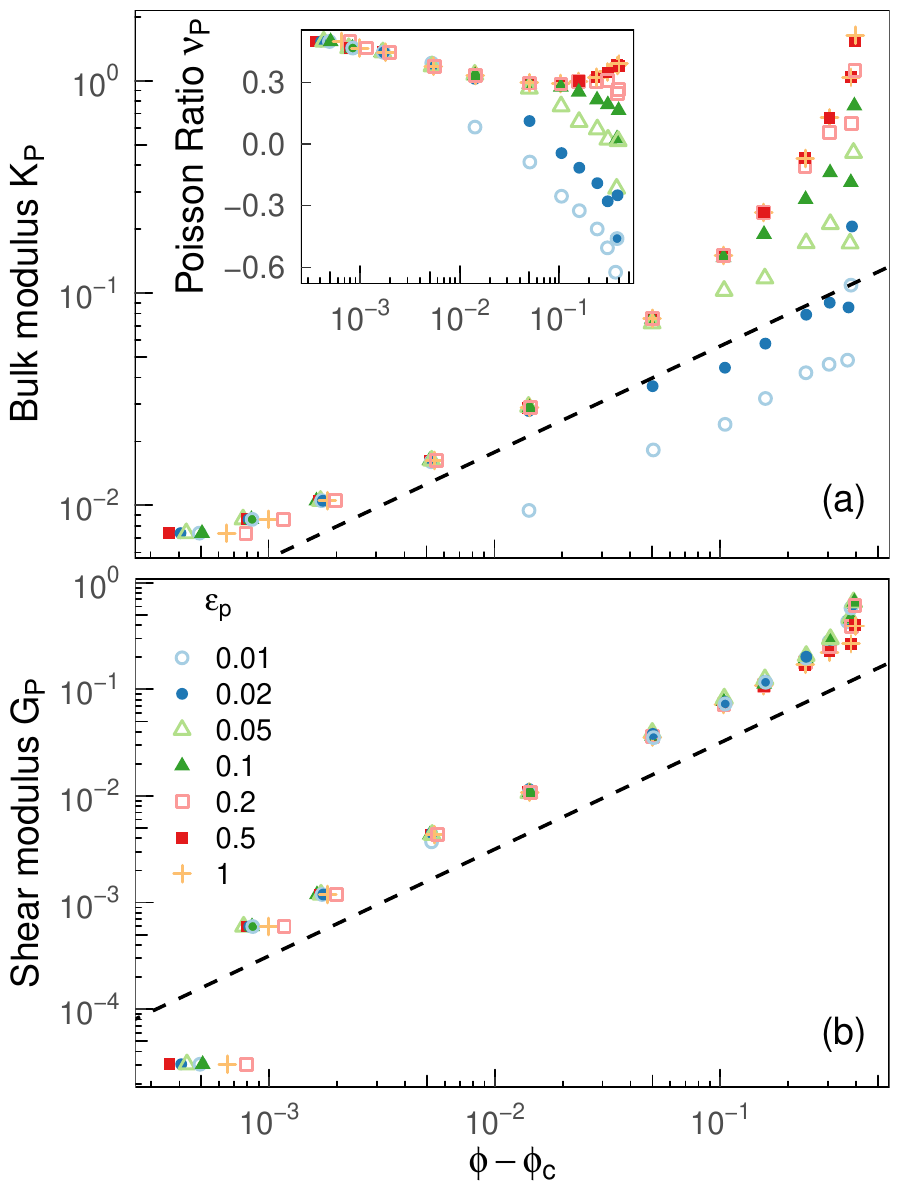}
	\caption{The (a) bulk modulus $K_P$ and (b) shear modulus $G_P$ of systems under confinement as a function of $\phi - \phi_c$ for different $\epsilon_p$ indicated in panel (b). Dashed lines are power laws with exponents (a) $1/2$ and (b) unity. The inset in panel (a) includes the corresponding Poisson's ratio $\nu_P$.}
	\label{fig:moduli}
\end{centering}
\end{figure}

During compaction, a related metric of interest is the current elastic properties of the packed grains.
Using the protocol described in greater detail in Ref. \cite{Clemmer2024}, systems are periodically replicated during compaction to precisely measure the bulk modulus $K_P$ (effectively the slope of Fig. \ref{fig:Pcurves}) and shear modulus $G_P$ of a confined packing.
This was done by applying small additional deformations, compacting 0.06\% strain for $K_P$ or shearing the box by 0.04\% strain for $G_P$, and fitting the change in the stress tensor with a least-mean-square regression.
In Fig. \ref{fig:moduli}(a), $K_P$ generally increases with $\phi$, rising approximately as the expected power law, $(\phi-\phi_c)^{1/2}$, near jamming \cite{OHern2003, Agnolin2007b}.
Mirroring the observed behavior in Fig. \ref{fig:Pcurves}, curves splay at an $\epsilon_p$-dependent packing fraction due to plastic softening which can result in a factor of ten difference between moduli near $\phi = 1.0$.

In the elastic limit, $G_P$ increases approximately proportional to $(\phi-\phi_c)$ as the system compacts, which is also expected near jamming \cite{OHern2003, Agnolin2007b, Wang2021}.
However, in contrast to $K_P$, $G_P$ exhibits relatively little dependence on $\epsilon_p$ (Fig. \ref{fig:moduli}[b]) because shear deforms the system in a different direction.
Thus, many of the bonds that resist the shear deformation have not yet plastically yielded and are still elastic, implying $G_P$ is significantly less sensitive to $\epsilon_p$.
Interestingly, this dichotomy has an unusual implication for the Poisson's ratio of the packed sample, $\nu_P = (3 K_P - 2 G_P) / (6 K_P + 2 G_P)$.
In the elastic limit with $\epsilon_p = 1.0$, $\nu_P$ first decreases then increases as $\phi$ increases, reaching a minimum around $\phi-\phi_c \sim 0.1$.
Yet in systems with plasticity, $\nu_P$ continues to drop and can reach well into the auxetic limit.
While this is unusual, it is important to note that these results may not be equivalent to what moduli would be measured if systems were first unloaded to zero pressure before being deformed. 
Regardless, it suggests plasticity in highly pressured granular material may lead to unexpected mechanical behavior.
In experimental work by \citet{Carnavas1998} and \citet{Hentschel2007} on powders that are likely ductile, including metals, the Poisson's ratio is found to reach a minimum, much like our elastic simulations, but does not exhibit auxetic behavior.
Exploring the differences between these systems and unraveling the underlying physics of this behavior remains an unexplored task.

\subsection{Onset of plasticity}
\label{sec:results_deviation}

\begin{figure}
\begin{centering}
	\includegraphics[width=0.9\columnwidth]{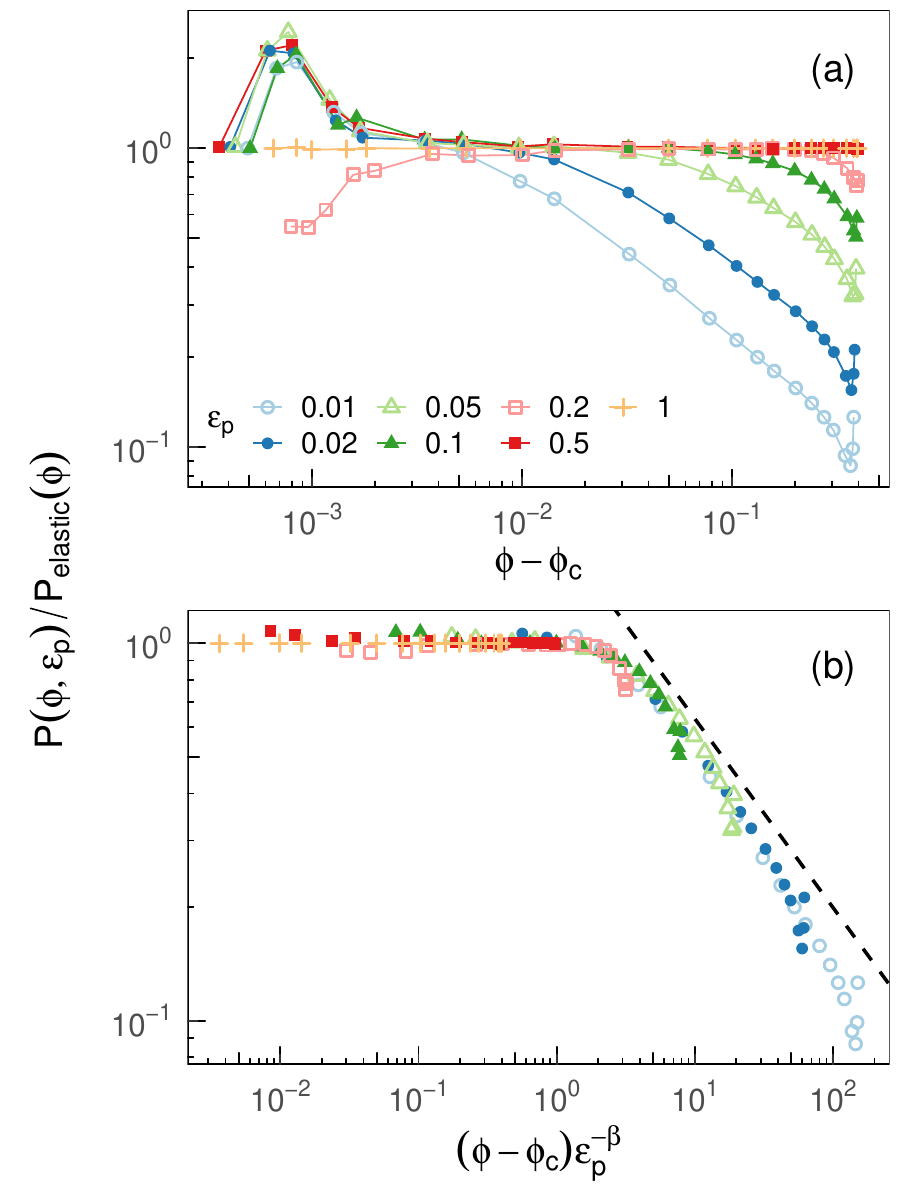}
	\caption{(a) Pressure normalized by the elastic limit as a function of $\phi - \phi_c$ at the indicated $\epsilon_p$. (b) Normalized pressure for $\phi-\phi_c > 2 \times 10^{-3}$ as a function of a scaled packing fraction with exponent $\beta = 1.3$. The dashed line represents a power law with exponent $-1/2$.}
	\label{fig:Pnormalized}
\end{centering}
\end{figure}

While Fig. \ref{fig:Pcurves} clearly demonstrates that plasticity has a large impact on compaction, it is harder to visually evaluate its magnitude and identify when plastic systems first deviate from the elastic limit.
To highlight the shift from the elastic limit, we therefore normalize pressure curves $P(\phi, \epsilon_p)$ by results in the elastic limit $P_\mathrm{elastic}(\phi)$.
Here $P_\mathrm{elastic}(\phi)$ is approximated by fitting a linear spline function to $P(\phi, 1.0)$ in Fig. \ref{fig:Pcurves}. 
By definition, $P(\phi, 1.0) / P_\mathrm{elastic}(\phi) \approx 1.0$ as seen in Fig. \ref{fig:Pnormalized}(a) where minute differences only arise from minor interpolation errors. 
At $\epsilon_p = 0.5$, the system is still practically indistinguishable from the elastic limit, however, interpolation errors lead to a noticeable bump at $\phi -\phi_c \sim 10^{-3}$ where the strong sensitivity to $\phi$ near jamming aggravates interpolation errors.
Therefore, subsequent discussion ignores comparisons at $\phi -\phi_c < 2 \times 10^{-3}$.

At $\epsilon_p < 0.5$, differences are richer.
Initially, the normalized pressure is constant before it suddenly drops with increasing $\phi$. 
The location of this transition, labeled $\phi^*(\epsilon_p)$, shifts lower as $\epsilon_p$ decreases and plasticity emerges at lower pressures.
Unexpectedly, scaling the excess packing fraction above jamming or $(\phi - \phi_c)$ by a power of $\epsilon_p$ with a nontrivial exponent of $\beta \sim 1.3$ collapses these curves in Fig. \ref{fig:Pnormalized}(b).
This suggests the transition can be described by $\phi^*(\epsilon_p) \sim \phi_c + \epsilon_p^\beta$.
Outside of the extreme limit of $\phi \approx 1$, the quality of this collapse is surprisingly good across the nearly two decades of $\epsilon_p$.
At $\phi > \phi^*(\epsilon_p)$, all curves approximately decay as a power of $(\phi - \phi_c$) with exponent $\sim 1/2$ over a domain that grows with decreasing $\epsilon_p$.

While the observed behavior is consistent with a power law, it is important to note that the domain only extends one to two decades so it is difficult to confirm its validity and we do not currently have a theoretical reason to expect it.
These observations would therefore benefit from simulations closer to the jamming transition, possibly with either more grains to remove finite size effects \cite{Goodrich2012} or a more rigorous jamming protocol \cite{Agnolin2007}, at smaller values of $\epsilon_p$ to extend the potential power-law domain.
However, the existence of a relatively simple relation, power law or not, between the onset of plasticity in the local bond model and within the macroscopic granular packing and a collapse of pressure curves presents an intriguing possibility for developing analytic models of granular compaction \cite{Kenkre1996, Storakers1999, Nezamabadi2021, Nezamabadi2021b}.

\subsection{Microscopic compaction response}
\label{sec:results_micro}

\begin{figure}
\begin{centering}
	\includegraphics[width=0.9\columnwidth]{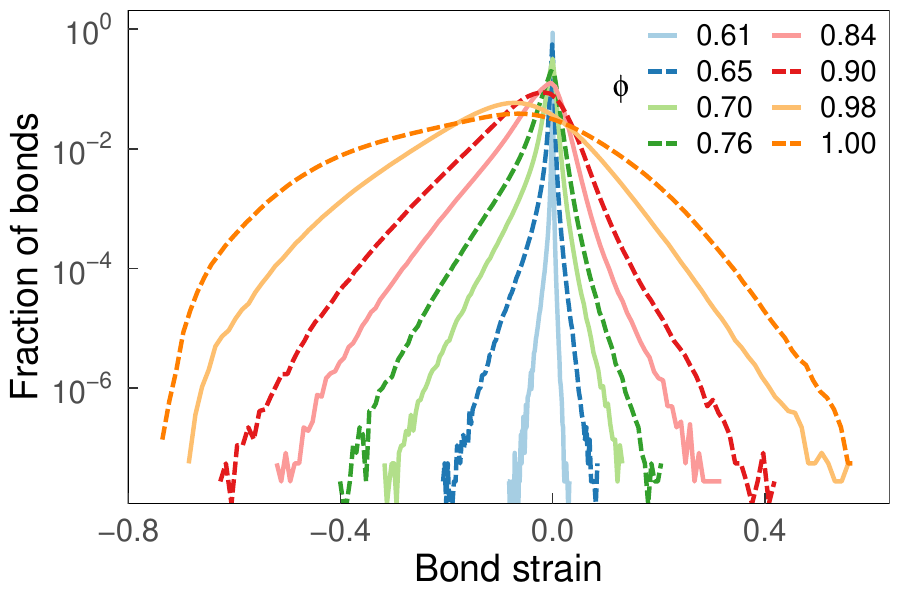}
	\caption{Fraction of bonds of a given strain in the elastic limit, $\epsilon_p = 1.0$, at the indicated packing fractions $\phi$ rounded to two digits.}
	\label{fig:elasticdist}
\end{centering}	
\end{figure}

In addition to extracting macroscopic properties, we also track particle and bond dynamics to gain insight into the microscopic response of the system to high pressures.
A particularly straightforward yet informative quantity is the local strain of bonds $\epsilon$ where we, again, first focus on the elastic limit.
Relatively close to jamming, $\phi \sim 0.61$ or $\phi-\phi_c \sim 0.01$, the distribution of bond strains is sharply peaked at zero with asymmetric tails (Fig. \ref{fig:elasticdist}).
More bonds are in compression than tension.
As the peak decays, there appears to be a slight kink in the distribution beyond which its tails look exponential.
Physically, this distribution reflects the fact that most of the elastic strain in a grain is isolated to relatively few bonds near the surface and adjacent to contacts.

As the system continues to compact and bonds stretch further, these general characteristics persist although the location of the kink shifts to strains of larger magnitude and becomes harder to identify as the exponential tails decay more slowly.
Remarkably, experimental studies by \citet{Vu2019} found qualitatively similar distributions of the internal von Mises strain of grains when compressing quasi-2D packings of hyperelastic cylinders.
Distributions were sharply peaked with asymmetric tails that broadened with increasing packing fraction.
Visually, there appears to be a similar kink, beyond which distributions resemble exponential decays.
Although the experiments did not extend beyond $\phi \approx 0.9$, in simulations we find this trend eventually breaks down when the system nears a fully dense state, $\phi \ge 0.98$, as the compressive tail begins to form a plateau.

\begin{figure*}
\begin{centering}
	\includegraphics[width=0.9\textwidth]{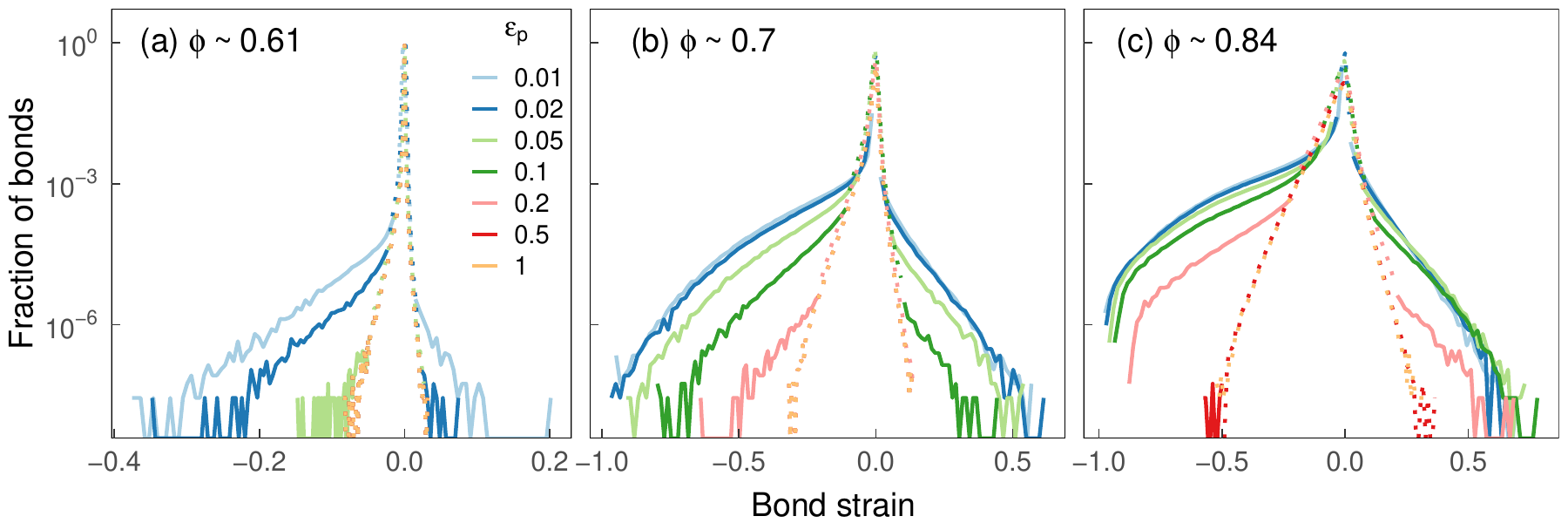}
	\caption{Fraction of bonds of a given strain $\epsilon$ in systems at packing fractions $\phi$ of (a) 0.61, (b) 0.7, and (c) 0.84 across values of $\epsilon_p$ listed in panel (a). For each series, data for $|\epsilon| < \epsilon_p$ is indicated by dotted lines while data for $|\epsilon| \ge \epsilon_p$ is indicated by solid lines.}
	\label{fig:distributions}
\end{centering}	
\end{figure*}

In systems with plasticity, the statistics of bonds is expected to be distorted at strains $\epsilon$ greater than $\epsilon_p$ due to the blunting of forces.
Close to jamming, $(\phi-\phi_c) \sim 0.01$ or $\phi \sim 0.61$, no discernible difference in distributions is detectable in systems with $\epsilon_p > 0.05$ which all conform to the fully-elastic limit (Fig. \ref{fig:distributions}[a]). 
This is simply due to the fact that no bonds have reached the plastic threshold.
In systems with $\epsilon_p \leq 0.05$, distributions exhibit a distinct departure from the elastic limit beginning at $\epsilon \sim \epsilon_p$ with a large excess of highly strained bonds.
The tail still appears exponential but with a slower $\epsilon_p$-dependent decay rate.
As $\phi$ increases (Fig. \ref{fig:distributions}[b] and [c]), distributions at higher $\epsilon_p$ undergo a similar departure from the elastic limit causing more weight to be placed in the plastic regime, $\epsilon > \epsilon_p$.
This eventually distorts the shape of the peak.
\citet{Vu2019} also measured von Mises strain distributions in experiments involving plastic material. 
While their measured distributions appeared functionally similar to their elastic material at low packing fractions, distinct evolution was also found with increasing compaction leading to a relative excess of large compressive strains 

\begin{figure}
\begin{centering}
	\includegraphics[width=0.9\columnwidth]{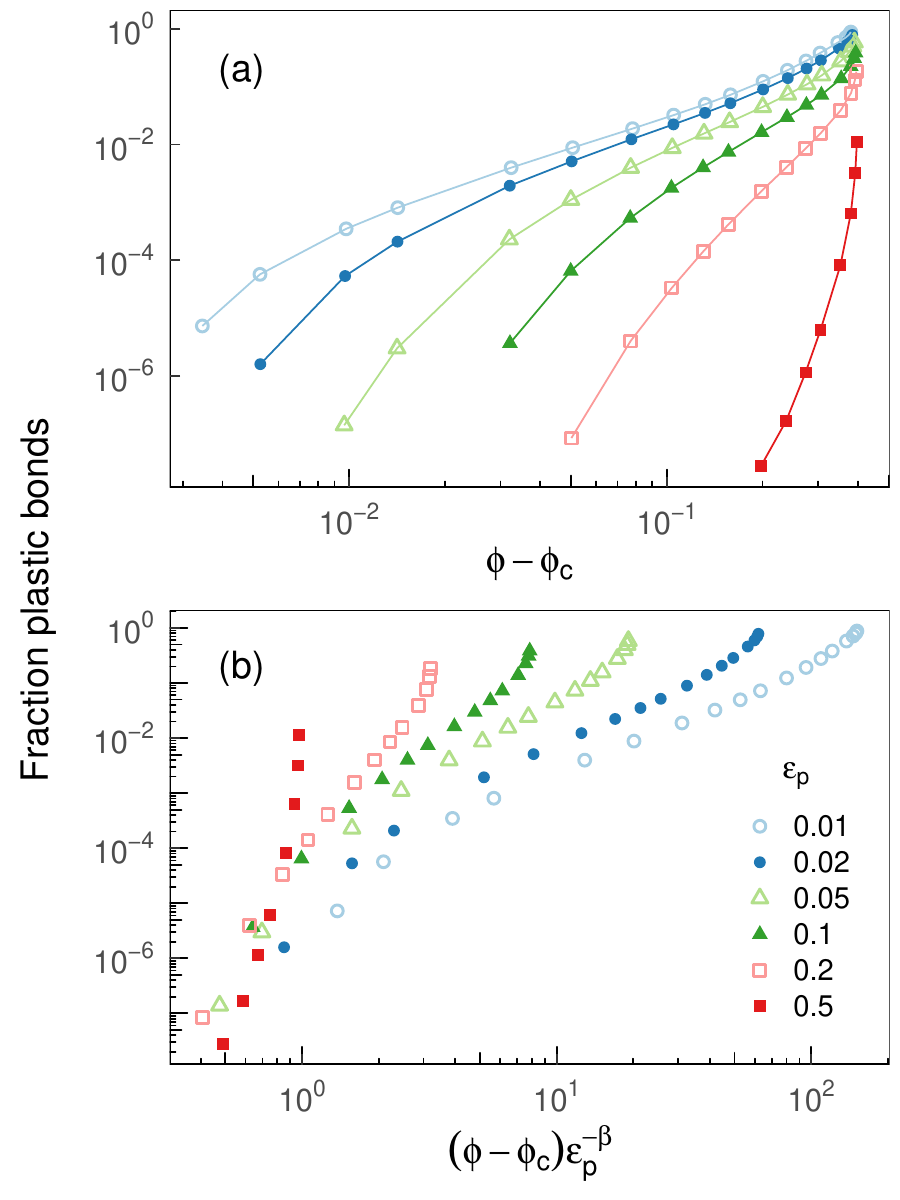}
	\caption{(a) The fraction of plastically activated bond as a function of excess packing fraction $\phi-\phi_c$ at various values of $\epsilon_p$. (b) The same data in panel (a) is plotted after scaling $\phi-\phi_c$ by $\epsilon_p^{-\beta}$}
	\label{fig:fplastic}
\end{centering}
\end{figure}

This stark departure from the elastic limit naturally suggests that the deviation in macroscopic behavior simply coincides with the first activation of plasticity in bonds.
To confirm this, we calculate the fraction of bonds that are plastically activate as a function of $(\phi-\phi_c)$, as seen in Fig. \ref{fig:fplastic}[a].
In the elastic limit, $\epsilon_p = 1.0$, this metric is always zero.
At $\epsilon_p < 1.0$, there is a sudden jump at a $\epsilon_p$-dependent value of $\phi$ as the number of plastically activated bonds rapidly grows.
Scaling the excess packing fraction by the previously observed factor of $\epsilon_p^{-\beta}$, we find that this initial jump does roughly coalesce at single point (Fig. \ref{fig:fplastic}[b]), but possibly at a smaller value than can be seen in Fig. \ref{fig:Pnormalized}.
This is consistent with the idea that the observed macroscopic scaling emerges from the extreme statistics of distributions.
While this may seem obvious since simulations first numerically deviate when the first bond plastically yields, the effect is quite dramatic given the exponential nature of the distribution tails and the relative rarity of highly strained bonds.

\section{Summary}
\label{sec:summary}

Using a conceptually simple bonded particle model of elastic-plastic granular material, we systematically explored the effects of a plastic yield strain $\epsilon_p$ on compaction.
With increasing packing fraction $\phi$, a system's pressure followed the elastic limit up until a threshold $\phi^*(\epsilon_p)$ that grew with increasing $\epsilon_p$ above which the system softened.
Isolating the transition, we found the threshold's dependence on $\epsilon_p$ is well described by a power law over nearly two decades, $\phi^*(\epsilon_p) - \phi_c \sim \epsilon_p^\beta$ where $\phi_c$ is the packing fraction at jamming and $\beta \sim 1.3$. 
After scaling the excess packing fraction by $\epsilon_p^{-\beta}$, we also found the relative decrease in pressure above this threshold collapsed across $\epsilon_p$.
These results provide a relatively simple functional description of the role of plasticity and, with more investigation, could serve as a basis for analytic predictive models \cite{Kenkre1996, Storakers1999, Nezamabadi2021, Nezamabadi2021b}.

To tease out the microscopic origin of this behavior, we considered the distribution of local bond strains.
Distributions were found to qualitatively reproduce observations from experiments of elastic and plastic granular materials by \citet{Vu2019} and our control over $\epsilon_p$ highlighted the emergence of of plasticity in the tails of distributions which generated excess weight relative to the elastic limit.
We found the extreme statistics, quantified by reducing the distribution to the fraction of bonds with strains greater than $\epsilon_p$, obeyed the same scaling with $\epsilon_p$ providing a connection between microscopic and macroscopic behavior.

While this work has begun to tease out fundamental relations between the plastic yield strain of a material and the onset and magnitude of softening in a compacted granular system, we do not have a theoretical explanation for the observations and, therefore, hope these findings stimulate further investigation into the topic.
This study was limited to an idealized material model with a simple bond construction.
It is unknown whether the observed relations are representative of all elastic-plastic materials or whether they may depend on the exact constitutive law.
Futhermore, this work only considered a simple loading protocol and avoided much of the complexity that is relevant to industrial processes. 
For instance, what happens after such systems are unloaded?

\section{Acknowledgments}

This work is funded by the Advanced Simulation and Computing program.
Sandia National Laboratories is a multi-mission laboratory managed and operated by National Technology \& Engineering Solutions of Sandia, LLC (NTESS), a wholly owned subsidiary of Honeywell International Inc., for the U.S. Department of Energy’s National Nuclear Security Administration (DOE/NNSA) under contract DE-NA0003525. This written work is authored by an employee of NTESS. The employee, not NTESS, owns the right, title and interest in and to the written work and is responsible for its contents. Any subjective views or opinions that might be expressed in the written work do not necessarily represent the views of the U.S. Government. The publisher acknowledges that the U.S. Government retains a non-exclusive, paid-up, irrevocable, world-wide license to publish or reproduce the published form of this written work or allow others to do so, for U.S. Government purposes. The DOE will provide public access to results of federally sponsored research in accordance with the DOE Public Access Plan.


\appendix
\section{Compaction of bulk solid}
\label{sec:bulk}

\begin{figure}
\begin{centering}
	\includegraphics[width=0.9\columnwidth]{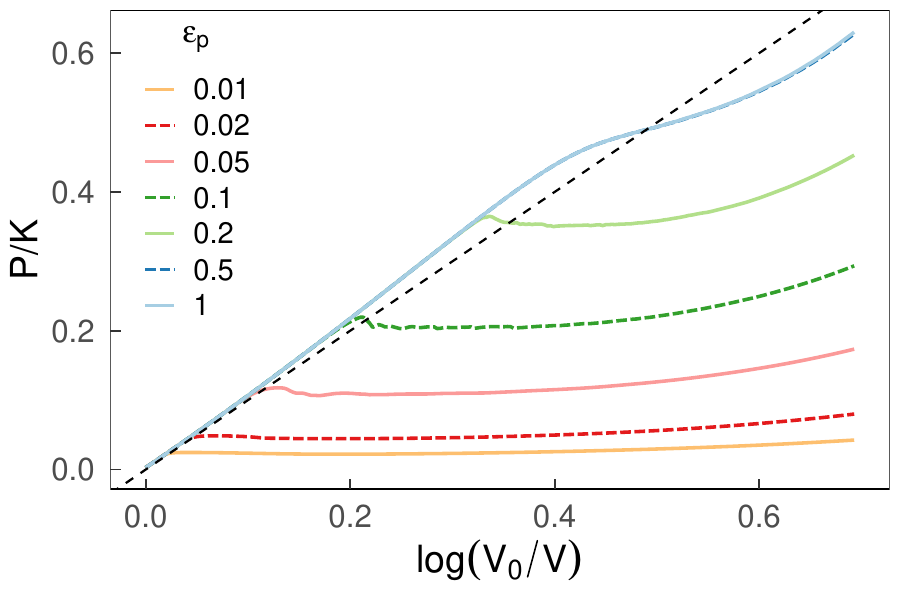}
	\caption{The pressure normalized by the bulk modulus $K$ as a function of true (logarithmic) volumetric strain for the indicated values of $\epsilon_p$. The dashed, black line has a slope of unity for comparison.}
	\label{fig:bulk}
\end{centering}
\end{figure}

While harmonic bonds produce an isotropic linear elastic response at small strains, as the strain increases there are growing deviations from linear elasticity \cite{Clemmer2024}.
This is due to the underlying glassy structure and non-affine displacement of particles.
To quantify this effect, bulk cubic systems representing a uniform solid (no grains) with side lengths of $50 d$ and fully periodic boundary conditions are isotropically compacted.
Compaction was performed at a constant true strain rate of $10^{-6} \tau^{-1}$.
Further simulations at a rate of $10^{-5} \tau^{-1}$ exhibited virtually no difference, suggesting there are minimal finite-rate effects. 
In Fig. \ref{fig:bulk}, the resulting pressure is plotted as a function of true strain for different values of $\epsilon_p$. 

In a system with purely elastic bonds, represented by $\epsilon_p \ge 0.5$, the pressure initially grows in proportion to the strain with a slope equal to the bulk modulus $K \approx 1.24 k/d$, measured in Ref. \cite{Clemmer2024}.
At strains above $\sim 10\%$, the pressure visibly exceeds the initial linear extrapolation, however, this effect is relatively small and might be captured by an approximately 10$\%$ increase in the slope.
Significant deviations only emerge around strains of $\sim 40\%$ at which curves are obviously non-linear. 
This instance corresponds to a large increase in the non-affine displacement of particles\footnote{The non-affine displacement of a particle is calculated by integrating its velocity across time, excluding the displacement due to the affine shift applied to particles' positions from the deformation protocol.} as the average magnitude of the non-affine displacement roughly doubles between strains of 40\% to 50\%.
This indicates that the particle/bond discretization of linear elasticity is beginning to break down at these high strains.
In granular systems considered in the main text, the average volumetric strain within grains only reaches comparable values at packing fractions $>99\%$ implying such effects may only emerge in one or two of the densest data points in Fig. \ref{fig:Pcurves}.

At lower values of $\epsilon_p$, plastic deformation softens the compaction of the bulk system.
The rise in pressure initially conforms to the elastic limit up to some threshold above which the system plastically deforms and the pressure nearly plateaus.
As $\epsilon_p$ decreases, this transition occurs at smaller strains as bonds plastically activate earlier.
 
\section{Additional bond statistics}
\label{sec:moments}

To describe the transition in compaction curves seen in Fig. \ref{fig:Pnormalized}, one could imagine there is a characteristic bond strain that grows as $\sim (\phi-\phi_c)^{1/\beta}$ and designates the onset of plastic softening when it exceeds $\epsilon_p$.
To extract such a characteristic strain from the distributions of bond strains $\epsilon$, one might consider many scalar metrics. 
In this appendix, we explore the evolution in a few common metrics and compare them to this hypothetical $1/\beta$ power law with $\beta = 1.3$.

\begin{figure}
\begin{centering}
	\includegraphics[width=0.9\columnwidth]{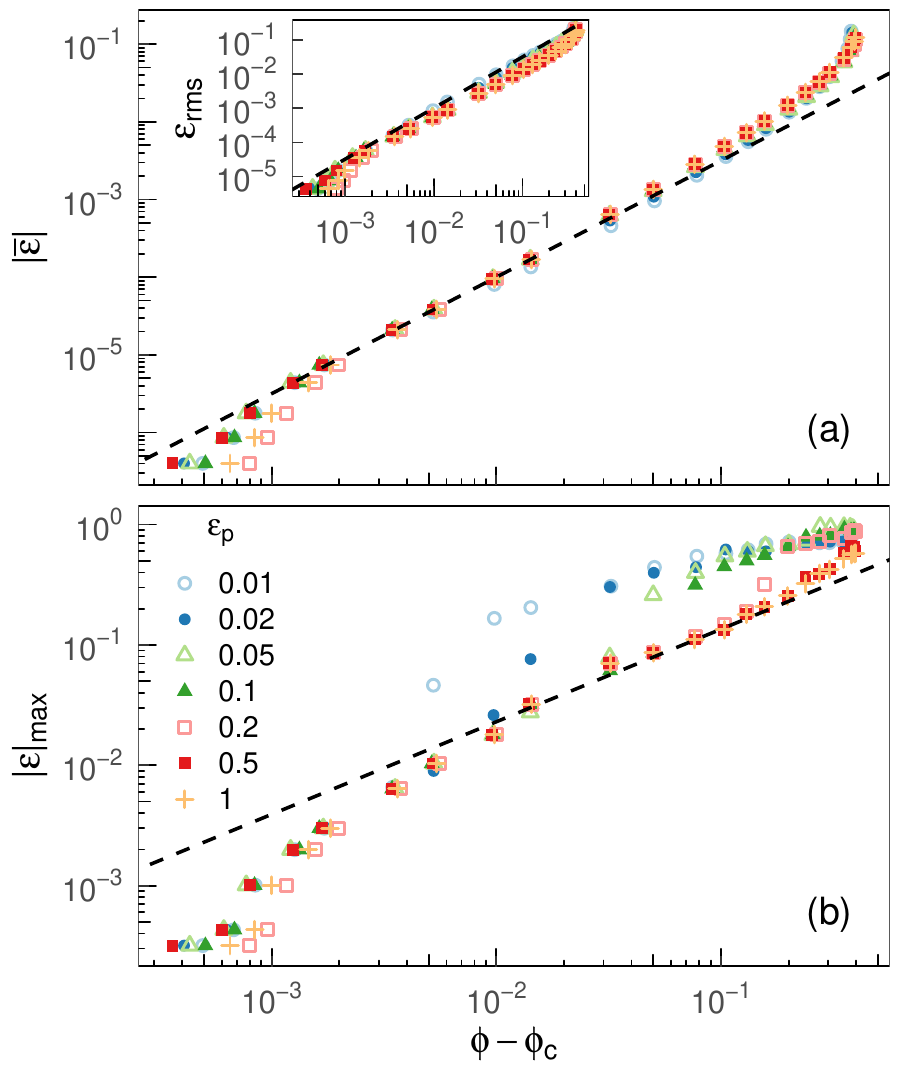}
	\caption{(a) Absolute value of the average bond strain, (inset) root mean square bond strain, and (b) maximum absolute bond strain as a function of $\phi-\phi_c$ for the indicated values of $\epsilon_p$. Dashed lines are power laws with exponents of $3/2$ in panel (a) and its inset and $1/\beta \approx 0.77$ in panel (b).}
	\label{fig:moments}
\end{centering}
\end{figure}

The first moment, or the mean $\bar{\epsilon}$, has little dependence on $\epsilon_p$ and grows as $\sim (\phi - \phi_c)^{3/2}$ (Fig. \ref{fig:moments}[a]).
This is essentially proportional to the pressure and is expected as the force of Hookean bonds is proportional to the strain. 
Similar behavior is seen both for the square root of the second moment (the root mean-square strain, in the inset) and the second cumulant (the standard deviation, not shown).
These low order metrics are not indicative of the plastic transition simply because they are dominated by the peaks of distributions while, as noted in the main text, the emergence of plasticity occurs at the tails.

While going to third order and higher moments may start to reveal some dependence on $\epsilon_p$ as they becomes more sensitive to the tail, instead we simply consider the maximum. 
While statistically noisier, we see in Fig. \ref{fig:moments}(b) that the maximum absolute bond strain does somewhat follow the expected $1/\beta$ power-law growth over approximately two decades in the elastic limit. 
The maximum also depends significantly on $\epsilon_p$, rapidly rising as bonds start to plastically activate.
Although not comprehensive, this analysis is consistent with the argument in the main text, emphasizing that the extreme statistics of the distributions are essential to quantifying the effect of plastic deformation on compaction.


\newpage
\bibliographystyle{elsarticle-num-names}

\newpage

\end{document}